\documentclass[aps,showpacs,twocolumn,superscriptaddress,prb]{revtex4-1}
\usepackage{graphicx,color,epsfig,rotate}
\usepackage{amssymb,amsmath,bm}
\begin{document}

\title{Field-dependence of magnon decay in yttrium iron garnet thin films}
\author{A. L. Chernyshev}
\affiliation{Department of Physics, University of California, Irvine,
California 92697, USA}
\date{\today}
\begin{abstract}
We discuss threshold field-dependence of the decay rate of the 
uniform magnon mode in yttrium iron garnet (YIG) thin films. 
We demonstrate that decays must cease to exist in YIG films of thickness less than  $1\,\mu$m,
the lengthscale defined by the exchange length. 
We show that due to the symmetry of the three-magnon coupling
 the decay rate is linear in $\Delta H\!=(H_c\!-\!H)$ in the vicinity of  the threshold field $H_c$
instead of the step-like 
$\Gamma_{{\bf k}=0}\!\propto\!\Theta\left(\Delta H\right)$ 
expected from the two-dimensional character of magnon excitations in such films. 
For thicker films, the decay rate should exhibit multiple  steps 
due to thresholds for decays into a sequence of the two-dimensional magnon bands.
For yet thicker films, such thresholds  merge 
and crossover to the three-dimensional single-mode behavior: $\Gamma_{{\bf k}=0}\propto |\Delta H|^{3/2}$.
\end{abstract}
\maketitle

{\it Introduction.}---Extensive experimental and theoretical research on the
ferromagnetic insulator, yttrium iron garnet 
[Y$_3$Fe$_2$(FeO$_4$)$_3$ or YIG]   was started more than
half a century ago has benefited from this material's exceptional purity, high 
Curie temperature, and relative simplicity of the low-energy magnon spectrum.\cite{saga} Recent 
discovery of the Bose-Einstein condensation of the highly occupied magnon states
created by microwave pumping in YIG thin films has attracted substantial interest.\cite{Demo_BEC}
 Recently, 
threshold effects due to the so-called three-magnon splitting 
have been reported  in a quasi-two-dimensional (2D) thin films of YIG under microwave  pumping 
and as a function of external field.\cite{Demo_11} 
In this, as well as in the other recent works,\cite{PtYIG} control over the  spin current 
enhancement in layered metal-ferromagnet structures by the three-magnon processes is sought. 
This calls for a deeper theoretical insight into the decay dynamics of such structures. 
Fundamentally, given its outstanding properties,\cite{saga} YIG
may offer a fertile playground in the studies of threshold phenomena, because
the decay conditions in it can be varied continuously 
by both film thickness and external magnetic field.

{\it Decays.}---In this work, we discuss the decay rate of 
the uniform 
mode (${\bf k}\!=\!0$ magnon)
in an insulating ferromagnetic thin film as a function of external magnetic field and film thickness.  
In particular, 
using magnon dispersion in the 
lowest-mode approximation, we outline the ranges of the field and film thickness that 
favor decays in YIG. 
Specifically, we show that kinematic conditions for three-magnon decays cannot be met 
for the YIG films of thickness  $d_{\rm min}\!\approx\! 1\,\mu$m or less and for the fields 
$H_c^{\rm max}\!\approx\!600$ Oe or higher.
The upper  limit on the external field is defined solely by the 
magnetization of a ferromagnet, in agreement with 
 Ref.~\onlinecite{Demo_11}. Less obvious is
the existence of the 
limit on the film thickness, which is
fixed by another fundamental characteristics of a ferromagnet: its exchange length.
 The physical reason for that limit
is the decreasing role of the long-range dipolar interactions with the decrease 
of the film thickness. 
The presence of such a limit must be important for the control of relaxation and transfer
of the spin current in layered structures, which rely on the three-magnon processes
in YIG.\cite{PtYIG}

We find that
the threshold field-dependence of the decay rate for the ${\bf k}=0$ magnon  
near the threshold field $H_c$ is $\Gamma_{{\bf k}=0}\!\propto\! |\Delta H|\cdot\Theta(\Delta H)$, where 
$\Delta H\!=\!(H_c-H)$,
because the three-magnon interaction vanishes along the direction of the film magnetization, 
which is also precisely 
the ${\bf k}$-direction where magnon band minima are located. 
This leads to 
a reduction of the phase space for decays
and results in a 
vanishing decay rate near the threshold, contrary the na\"{i}ve expectation of the step-like
increase of $\Gamma_{{\bf k}=0}\!\propto\! \Theta(\Delta H)$, when the symmetry of the three-magnon
interaction is ignored.\cite{Rezende} 
We have supported our consideration by explicit calculations of both $T=0$ and 
$T=300$K relaxation rate dependencies on the field for several
representative YIG film thicknesses.

The finite-size quantization in the film thickness 
direction leads to the formation of multiple magnon bands.\cite{Kopietz} 
We argue that the field-dependence for the decay rate should exhibit multiple steps 
linear in $|H_{c_i}\!-\!H|$,  corresponding to the $H_{c_i}$ 
thresholds for decays into a sequence of magnon bands.
For thick films, these multiple thresholds should merge in a continuum, and the
decay rate will crossover to a three-dimensional (3D) single-band behavior:
 $\Gamma_{{\bf k}=0}\propto |\Delta H|^{3/2}$.
These results should be helpful in finding an optimal set of parameters for spin-current
enhancement.\cite{Demo_11}

\begin{figure}[t]
\includegraphics[width=0.99\columnwidth]{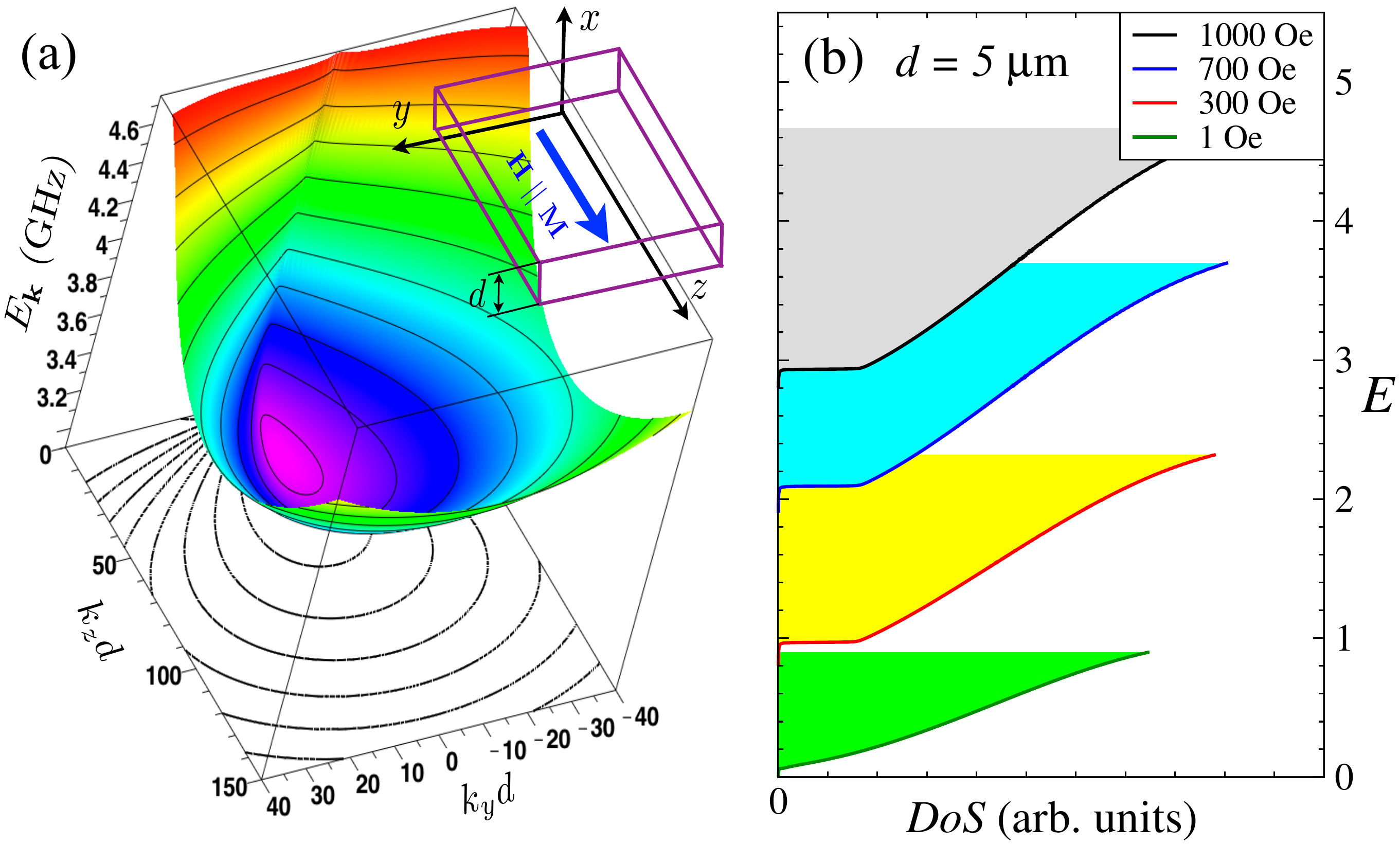}
\caption{(Color online) (a) 2D plot of magnon dispersion 
in YIG in the lowest-mode approximation, Eq.~(\ref{Ek}), 
for $H=1000$ Oe, $d=5$ $\mu$m, and in $k_z>0$ sector.
Inset: axes conventions relative to the film and the field/magnetization direction.
(b) 2D magnon density of states (DoS) for several representative field values, $E$ is in GHz.} 
\label{Fig1}
\end{figure}

{\it Dispersion, density of states.}---Despite its fairly complicated crystal structure, 
at low energies YIG can be described with great success as an effective large-spin 
Heisenberg ferromagnet on a cubic lattice with nearest-neighbor exchange, long-range
dipolar interactions, 
and negligible spin anisotropy.\cite{saga,Melkov} 
Thus, at long wavelength, magnon energy in YIG is determined by a 
competition between three  couplings: exchange, dipolar, and Zeeman.
For a ferromagnetic crystal of the film geometry and external field directed in-plane
where it co-aligns with the magnetization direction, as is done most commonly in experiments,
the lowest-mode approximation for the magnon energy yields:\cite{Slavin,Tupitsyn,Rezende,Kopietz}
\begin{eqnarray}
\label{Ek}
E_{\bf k}=\sqrt{\Big(h+\rho {\bf k}^2+\widetilde{\Delta}_{\bf k}\sin^2\theta_{\bf k}\Big)
\Big(h+\rho {\bf k}^2+\Delta_{\bf k}\Big)}  , 
\end{eqnarray} 
where  $\Delta_{\bf k}= f_{\bf k}\Delta$ and $\widetilde{\Delta}_{\bf k}=(1-f_{\bf k})\Delta$
and $h=\mu H$, $\rho =JSa^2$, and $\Delta=4\pi\mu M$ 
are the energy scale parametrizations of the external field, exchange, and dipolar interactions, respectively.
The form factor $f_{\bf k}=(1-e^{-|{\bf k}|d})/|{\bf k}|d$ is
from the  dipolar sums in the direction of the film thickness $d\gg a$, 
and $\theta_{\bf k}$ is the angle between the ferromagnet's magnetization (directed in-plane, $z$ axis)
and magnon's in-plane 2D wave vector ${\bf k}$, see Fig.~\ref{Fig1}(a). We use $\mu\!=\!g\mu_B$ where
$g=2$ is an effective $g$ factor and $\mu_B$ is the Bohr magneton.
In this work we adhere to the notations and units of Ref.~\onlinecite{Kopietz}, 
which has provided a detailed microscopic spin-wave theory of YIG in $1/S$ approximation,
and we use experimental parameters for YIG,  magnetization $4\pi M=1750$ G, exchange 
stiffness $\rho/\mu=5.17\cdot 10^{-13}$ Oe~m$^2$, and lattice constant $a=12.376$ \AA.
In Fig.~\ref{Fig1}(a),  the magnon dispersion from Eq.~(\ref{Ek}) for a representative field $H=1000$ Oe
and film thickness $d=5$ $\mu$m is shown.

It should be noted that the dipolar interactions are responsible for the  
nontrivial structure of magnon band with the minima at finite wave vectors, 
which allow  the decay conditions to occur.
Although approximate,  Eq.~(\ref{Ek}) provides a close 
quantitative description of the lowest 2D magnon energy band,\cite{Kopietz}
quantized due to finite film thickness $d$ 
in the $x$ direction [Fig.~\ref{Fig1}(a), inset].
At small ${\bf k}$, dipolar interactions dominate over the exchange and result 
in a steep decrease from the energy of the uniform mode, 
$E_0\!=\!\sqrt{h(h+\Delta)}$, for ${\bf k}$'s along the magnetization direction. 
At larger ${\bf k}$, exchange energy dominates, giving $E_{\bf k}\!\approx\! h\!+\!\rho {\bf k}^2$; 
while at intermediate ${\bf k}$, competition between exchange and dipolar terms
results in peculiarly shaped minima at $\pm{\bf k}_m\!=\!(0,\pm k_m)$, see Fig.~\ref{Fig1}(a). 

The 2D density of magnon states from Eq.~(\ref{Ek}) is shown in  
Fig.~\ref{Fig1}(b) for several representative field values.
Predictably, 2D DoS exhibits  a step-like increase at the
band minimum, which is  followed by an unusual, almost linear increase, reminiscent
of the similar behavior for the relativistic  dispersion. The latter  is  
due to a nonparabolicity of the long-wavelength magnon dispersion, see 
Fig.~\ref{Fig1}(a).

\begin{figure}[t]
\includegraphics[width=0.92\columnwidth]{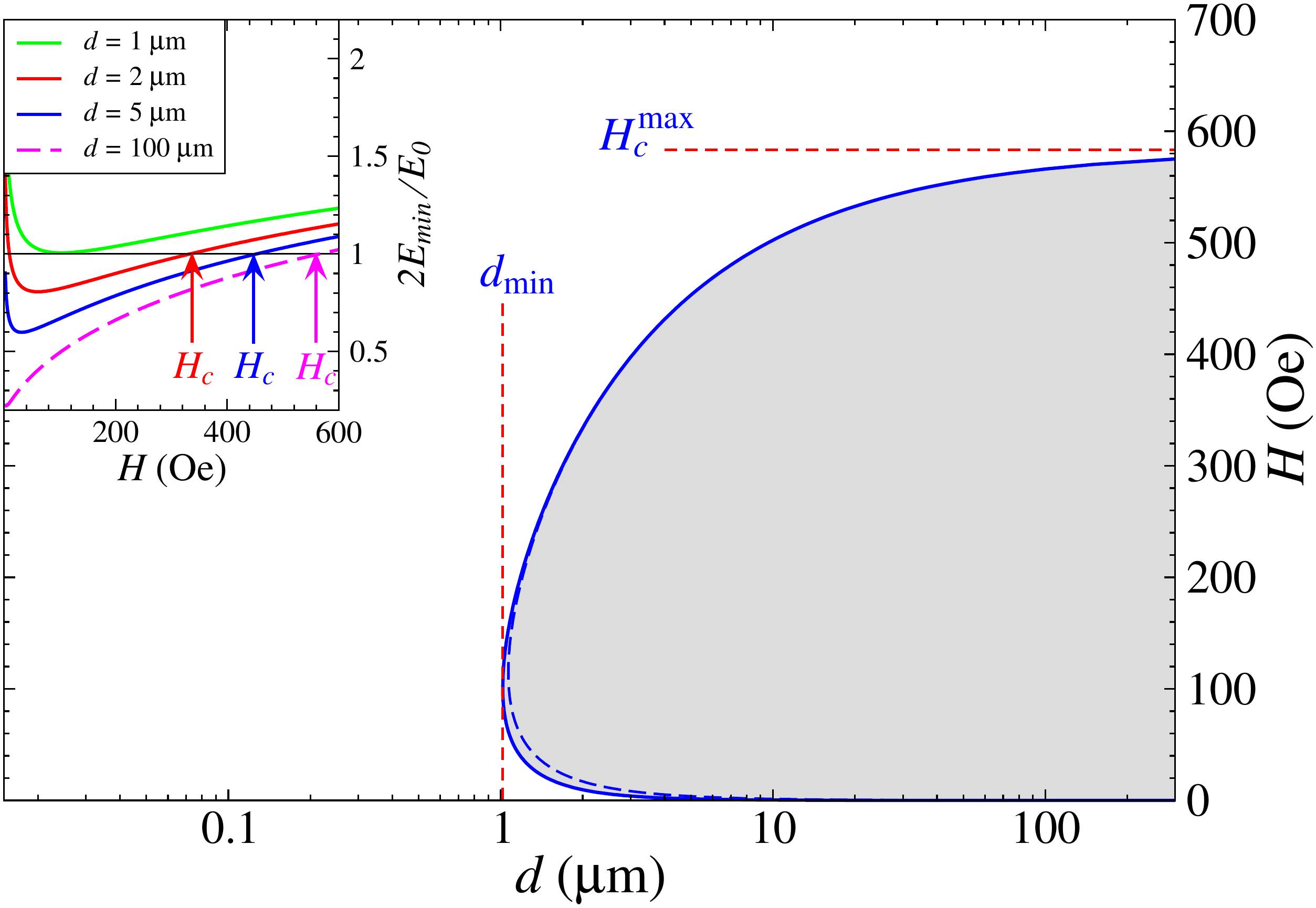}
\caption{(Color online) 
$d\!-\!H$ decay diagram for YIG, shaded area is
where decays are allowed. Solid (dashed) boundary is the numerical (analytical) 
solution for the threshold boundary, see text. $H_{\rm max}$ and $d_{\rm min}$ are shown. 
Inset: $2E_{\rm min}/E_0$ vs  $H$ for several $d$'s, upper threshold fields are indicated.} 
\label{Fig3}
\end{figure}

{\it ``Decay diagram''.}---For the decays to take place the kinematic conditions must be fulfilled. 
For the two-magnon decay (three-magnon splitting) 
of the uniform mode, the condition to be satisfied is simply $E_0=2E_{\bf q}$.
With the microscopic parameters, such as exchange stiffness and magnetization, fixed,
some other parameters can be varied to allow or  to forbid decays altogether.
As is clear from Eq.~(\ref{Ek}), the external field increases the energies of 
both the uniform mode and the minimum, making decays kinematically 
impossible at some higher field value.\cite{Demo_11}
Another parameter is the film thickness $d$, which enters Eq.~(\ref{Ek}) through the form factor
$f_{\bf k}$.  While the manner in which $d$ influences decays   is not {\it a priori} clear, both 
trends, vs field and vs thickness, can be examined numerically. Such an examination is 
exemplified in  Fig.~\ref{Fig3}, 
which shows $2E_{\rm min}/E_0$ vs field for several film thicknesses. Clearly, when the plotted quantity
is $<1$, decays are allowed, and the crossing of 1 corresponds to a threshold field for decays.

Our Fig.~\ref{Fig3} gives 
the complete $d\!-\!H$ ``decay diagram'' for YIG with the shaded area showing the parameter space 
where decays are allowed.  Two key results are clear in Fig.~\ref{Fig3}: 
(i) the upper threshold field does not  exceed some $H_c^{\rm max}$ even for large values of $d$, 
and (ii) there exists a lower limit on the film thickness
$d_{\rm min}$, below which the decays of the uniform mode may not occur at all. 
The solid boundary is the numerical solution of Eq.~(\ref{Ek}) for  the energy 
minimum and the decay conditions.
The dashed line is an approximate analytical solution, which turns out to be very precise. The latter
also gives us a  deeper insight into the nature of $H_c^{\rm max}$ and $d_{\rm min}$ discussed next.

{\it Large-$|{\bf k}_m|d$ approximation}---At large enough $d$ 
the wavevector of the magnon energy minimum 
satisfies $|{\bf k}_m|d\!\gg \!1$. Then the formfactor 
$f_{{\bf k}_m}\!\approx\! 1/|{\bf k}_m|d$ is small, reflecting the reduced role of dipolar interactions in the energy
of the ${\bf k}_m$-magnon. One can then show\cite{Supp} that both the exchange and the dipolar energies
for the  ${\bf k}_m$-magnon scale as $\propto\!d^{-2/3}$
and thus are $\ll h$ for any reasonable field. This implies that
the energy of the magnon band minimum is $E_{\rm min}\!\approx \! h$,  expected 
for the uniform mode in the absence of dipolar interactions. Then, the 
decay threshold equation, $2E_{\rm min}\!=\!E_0$,  trivially gives 
$h_c\!=\!\Delta/3$, relating saturated value of the decay threshold field 
to the material's magnetization:
$H_c^{\rm max}\!=\!4\pi M/3$ ($=583$ Oe for YIG in Fig.~\ref{Fig3}(b)), see also Ref.~\onlinecite{Demo_11}.

The physical question is: what parameter of the ferromagnet defines  $d_{\rm min}$?
One can extend the large-$|{\bf k}_m|d$ approach  and find\cite{Supp} 
that the energies of the dipolar and exchange interactions must be related by
$|{\bf k}_m|d\!=\!\left(\Delta d^2/4\rho\right)^{1/3}$.
With this, the decay threshold condition $2E_{\rm min}\!=\!E_0$
leads to an algebraic equation in $H_c$ vs $d$, which can be resolved in a compact form.\cite{Supp}
It is plotted as a dashed line in Fig.~\ref{Fig3}(b), which coincides almost exactly with the
decay boundary obtained from Eq.~(\ref{Ek}) numerically. From the same equation 
we find the minimal thickness to be $d_{\rm min}\!\approx \!C\sqrt{\rho/\Delta}$, 
explicitly related to the exchange length of the ferromagnet, 
$\ell_{ex}\!=\!\sqrt{\rho/\Delta}$, albeit with a large numerical coefficient 
$C
\approx 62.04$.\cite{Supp}
Using parameters for YIG, the exchange length is $\ell_{ex}\!\approx\! 13.9 a$ and
the  minimum thickness  $d_{\rm min}\!=\!1.067$ $\mu$m, remarkably close to the
numerical result $d_{\rm min}\!=\!1.017$ $\mu$m.

 The physical reason for the very existence of $d_{\rm min}$ 
is the decreasing role of long-range dipolar interactions in the magnon's energy 
with the decrease of film thickness, as the relation of $d_{\rm min}$ to exchange length implies.
The fact that the decay boundary in $d$ exceeds the exchange length by a large numerical factor 
is due to the rather stringent requirements  of the decay conditions.
We emphasize that the provided $d\!-\!H$ diagram should apply equally to the other thin-film ferromagnets.  

{\it Decay rate.}---Transitions that involve changing the number of magnons, 
such as decays, recombination, or coalescence, 
originate from the dipolar interactions that couple longitudinal and transverse spin components 
and therefore do not conserve
magnetic\cite{saga,Melkov} as well as mechanical angular momentum.\cite{Demo_11}
Microscopically, dipolar interactions result in anharmonic couplings of magnons, which,
for the decay processes of the ${\bf k}=0$ uniform mode into two magnons 
at ${\bf q}$ and $-{\bf q}$, can be written as
\begin{eqnarray}
\label{H3}
{\cal H}^{(3)}=\frac12\sum_{\bf q} V^{(3)}_{0;{\bf q}, -{\bf q}}  
\left(a_{\bf q}^\dag a_{\bf -q}^\dag a^{\phantom \dag}_{0} +{\rm H.c.} \right). 
\end{eqnarray} 
The three-magnon coupling in Eq.~(\ref{H3}) has an angular dependence: 
$V^{(3)}_{0;{\bf q}, -{\bf q}} =V_0\sin 2\theta_{\bf q}$, with $V_0=\Delta/\sqrt{2S}$.\cite{Sparks,Sparks1,saga}
 This angular
dependence is essential since it reflects the  symmetry of dipolar coupling of the transverse, $S^x$ ($S^y$), 
and longitudinal, $S^z$, spin components: $V^{xz}\propto xz/r^3$ ($V^{yz}\propto yz/r^3$).
In particular, it is natural for this coupling to vanish for the spin-wave propagating 
with the momentum ${\bf q}$  along the direction of magnetization ${\bf M}$ 
[$z$ axis, Fig.~\ref{Fig1}(a)].\cite{Sparks} We  would like to point out that it is also 
precisely the direction along which the minima of the magnon band are located.\cite{Kopietz1} 
Therefore, the
amplitude of the decay of ${\bf k}\!=\!0$ magnon into two  magnons at the band minima 
${\bf q}_{m}$ and $-{\bf q}_{m}$ is zero. 
This will lead to a rather spectacular violation of  the na\"{i}ve expectation: while kinematic 
 conditions for  the decay of ${\bf k}\!=\!0$ magnon into $\pm{\bf q}_{m}$ are just met 
 at $H_c$, the corresponding decay amplitude is vanishing. Thus, the decay rate must increase 
 gradually from the threshold, not in a jump-like fashion as in the DoS,  Fig.~\ref{Fig1}(b).

At $T=0$ only spontaneous magnon decays are allowed.\cite{review} 
The three-magnon recombination processes have to obey the same kinematic constraints as the 
decay, having therefore the same threshold conditions. The three-magnon coalescence processes 
involving the ${\bf k}\!=\!0$ mode correspond to the ``vertical'' transitions, which are forbidden 
either kinematically as in the single mode case or by the quantum number of the interband transition
in the multi-band situation. The four-magnon scattering amplitude from the exchange interaction
vanishes identically for the uniform mode, while the remaining four-magnon interactions from the 
dipole-dipole interaction together with impurity scattering should be providing 
a background with weak $H$ and $T$ dependence, distinct from the threshold behavior discussed
here. 
\begin{figure}[t]
\includegraphics[width=0.9\columnwidth]{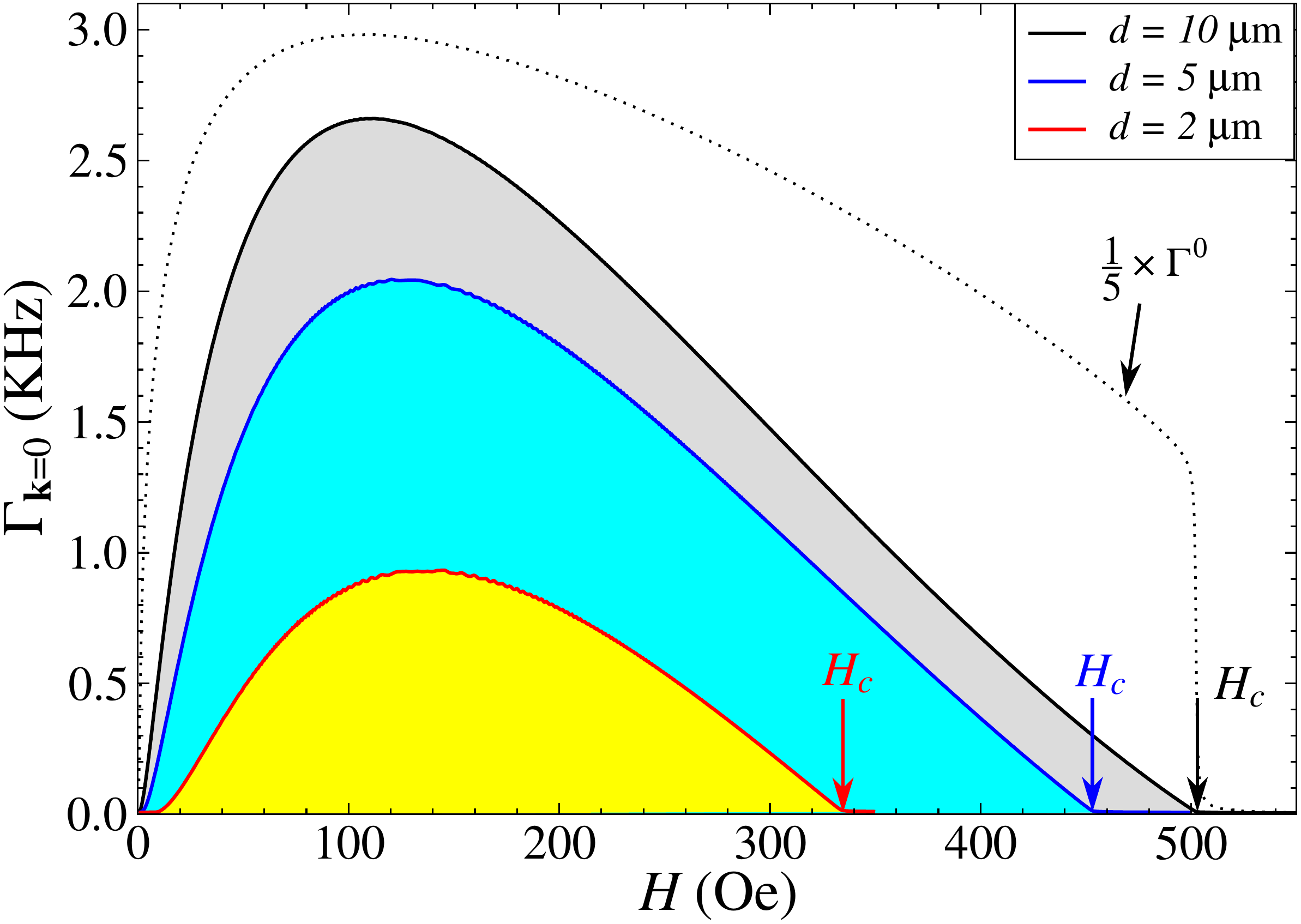}
\caption{(Color online) The $T=0$ decay rate $\Gamma_{{\bf k}=0}$ vs $H$ for $d=2$, 5, and 10 $\mu$m. 
Dotted line is  $\frac15\Gamma_{{\bf k}=0}$ for $d=10$ $\mu$m 
with the angular-dependence of the three-magnon coupling omitted, 
$V^{(3)}_{0;{\bf q}, -{\bf q}} \Rightarrow V_0$.} 
\label{Fig4_0}
\end{figure} 

With this in mind, using the kinetic approach,\cite{saga,Melkov} which takes into account the balance between
decay and recombination processes in the relaxation time approximation, the decay rate\cite{footnote} 
of the uniform mode is 
\begin{eqnarray}
\label{G}
\Gamma_{{\bf k}=0}=\pi\sum_{\bf q} \big|V^{(3)}_{0;{\bf q}, -{\bf q}}\big|^2\,\big[2n_{\bf q}+1\big]\,
\delta\left(E_0-2E_{\bf q}\right) 
\end{eqnarray} 
where $n_{\bf q}=[e^{\hbar E_{\bf q}/k_bT}-1]^{-1}$ is the Bose occupation factor.
It is clear from Eq.~(\ref{G}) that while the magnitude of the decay rate at finite $T$ can be
substantially modified from the $T=0$ result by the  Bose-occupation factors, the qualitative 
threshold behavior must remain the same.

\begin{figure}[t]
\includegraphics[width=0.9\columnwidth]{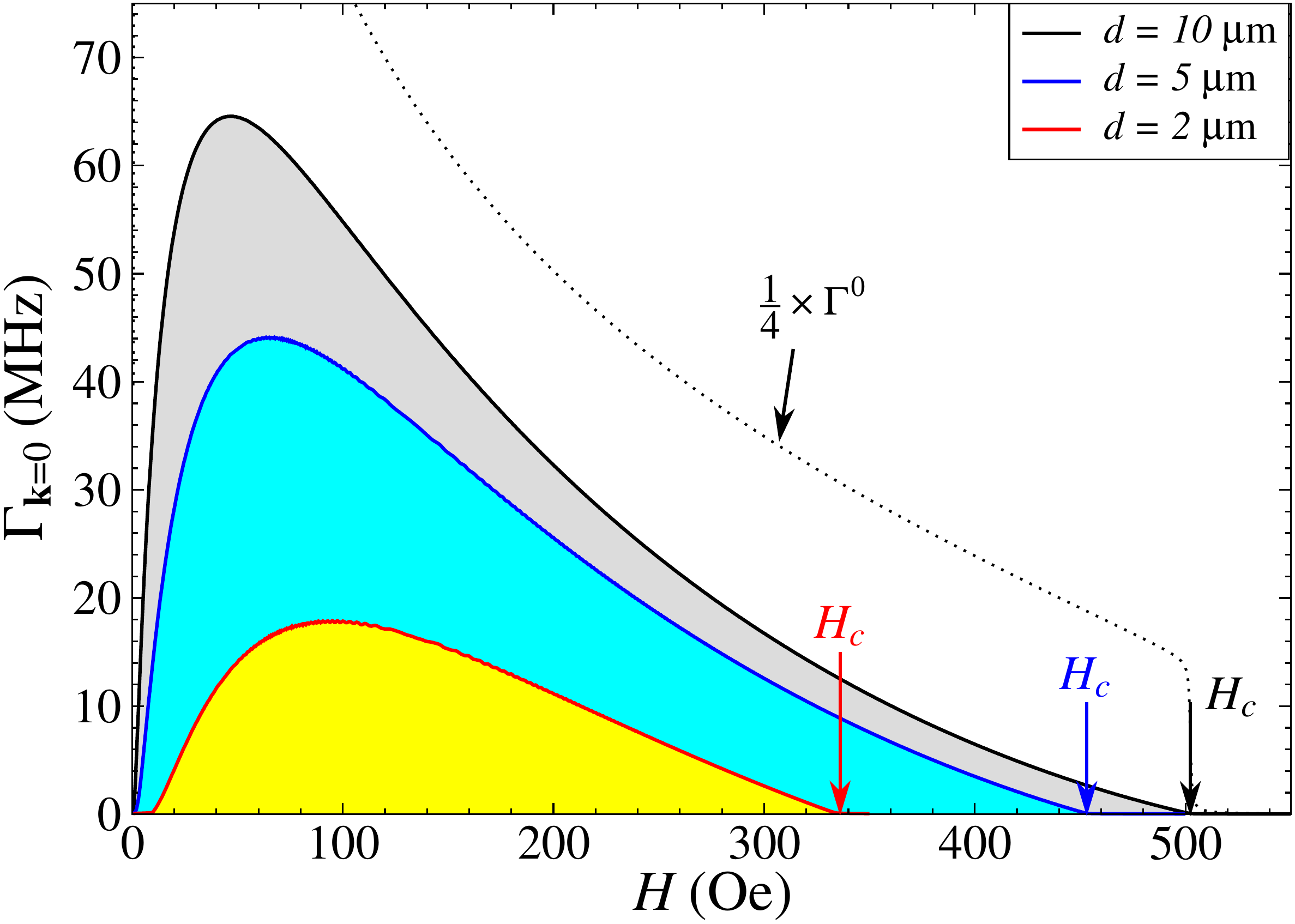}
\caption{(Color online) Same as in Fig.~\ref{Fig4_0}, 
but for $T=300$ K.} 
\label{Fig4}
\end{figure}

One can investigate the threshold behavior of Eq.~(\ref{G}) analytically
and obtain
$\Gamma_{{\bf k}=0}\!\propto\!|E_0-2E_{\rm min}|$  for $H\!\rightarrow\!H_c$.\cite{Supp}
Given the proportionality between $(E_0-2E_{\rm min})$ and  $\Delta H\!=\!(H_c\!-\!H)$ demonstrated in
Fig.~\ref{Fig3}, this yields
 {\it linear} dependence of the decay rate on the field relative to the threshold:
$\Gamma_{{\bf k}=0}\!\propto\! |\Delta H|\cdot\Theta(\Delta H)$, see Figs.~\ref{Fig4_0} and \ref{Fig4}.
Once again, this is the consequence of an effective suppression of the phase space for decays 
due to the discussed angular  dependence of the three-magnon coupling.

In Figs.~\ref{Fig4_0} and \ref{Fig4} we show $\Gamma_{{\bf k}=0}$ for three different film thicknesses
and for $T=0$ and $T=300$ K, respectively.
While the overall scale in the two figures is in a completely different frequency range,\cite{Supp} 
the shapes of $\Gamma$ vs $H$
are qualitatively  very similar,  especially concerning their threshold behavior vs field, in agreement
with the above analysis.
The relative difference between the curves for different $d$'s reflects
the smaller phase space for decays in thinner films. In the same Figs.~\ref{Fig4_0} and \ref{Fig4} 
we demonstrate 
a dramatic contrast with the results of  Eq.~(\ref{G}) if the angular dependence 
of the three-magnon coupling in Eq.~(\ref{H3}) is neglected, $V^{(3)}_{0;{\bf q}, -{\bf q}}\!\Rightarrow\!V_0$. 
The latter results exhibit jumps at $H_c$'s and a linear increase after that, similar 
to the 2D magnon DoS in Fig.~\ref{Fig1}(b). One should also note that the overall decay rate 
is also markedly  overestimated if the  angular dependence 
of the three-magnon interaction is ignored.

In thicker films, the decay rate will be further modified by 
the multiple magnon bands that occur due to finite-size quantization 
in the film thickness direction.\cite{Kopietz} It will exhibit a sequence of steps linear in $|H_{c_i}\! -\!H|$,
where $H_{c_i}$ is the threshold field for decay into an $i$th band,  increasing the decay rate 
every time the corresponding kinematic conditions are met.
Strictly speaking, the angle $\theta_{\bf q}$ in $V^{(3)}_{0;{\bf q}, -{\bf q}}$ is between the 3D ${\bf q}$ vector and 
magnetization vector ${\bf M}$. In the quasi-2D geometry with the levels quantized in the $x$ directions,
$q_x$ has discrete values and the minimal value of 
$\theta^{\rm min}_{{\bf q}^i_{\rm m}}\approx q^{\rm min}_{x,i}/|{\bf q}|$ is zero only for 
the lowest magnon band. Thus, the thresholds in $H_{c_i}$'s will have some small step-like behavior.
However, this effect must be negligibly smaller compared to the steps in Figs.~\ref{Fig4_0} and \ref{Fig4} 
when the angular dependence in $V^{(3)}$ is ignored altogether.
Extending our analysis to the limit of thicker films where multiple bands merge into a single 3D band, 
using Eq.~(\ref{G}) we obtain\cite{Supp} that the decay rate near the threshold field will crossover to 
$\Gamma_{{\bf k}=0}\!\propto\! |\Delta H|^{3/2}$.

{\it Conclusions.}--- In this work, we have discussed the 
field-dependence of the decay rate  of the 
uniform magnon mode in YIG thin films and the effects of film thickness in it.
As a result of our analysis, we have established that the two key characteristics of a ferromagnet,
its magnetization and exchange length, define the extent of the $d-H$ parameter range that favors decays 
in thin film geometry. Our calculations of the decay rate should 
provide an important guidance for the experimentalists in designing the optimal 
conditions for the control of spin current and its relaxation in thin films. A particularly intriguing suggestion to 
pursue is to study the properties of a thin film with varying thickness, which may permit decays and, as a
consequence, the spin current enhancement in one part of the film and forbid it in the other.

{\it Acknowledgments.}---I am deeply indebted to Doug Mills for introducing 
me to this problem, for many illuminating discussions and constant encouragement. 
I  thank Ilya Krivorotov for numerous enlightening conversations, his genuine 
interest and support, as well as generosity with his time spent on educating me in this field.
I also acknowledge useful conversation with Andreas Kreisel. This work was supported 
by the U.S. DoE under grant DE-FG02-04ER46174.


\section{Supplemental material}
\setcounter{equation}{0}


\subsection{large-$|{\bf k}_m|d$ approximation}

Assuming that for large enough $d$ the  magnon energy minimum 
satisfies $|{\bf k}_m|d\gg 1$, makes the formfactor 
$f_{\bf k}=(1-e^{-|{\bf k}_m|d})/|{\bf k}_m|d\approx 1/|{\bf k}_m|d$ small
and allows to simply $E_{\bf k}$ in  Eq.~(\ref{Ek}) 
to an algebraic expression in reduced energy scales of the 
dipolar and exchange interactions, $\bar{\Delta}=\Delta/|{\bf k}_m|d$ and 
$\bar{\rho}=\rho|{\bf k}_m|^2$:
\begin{eqnarray}
\label{Ekapp}
E_{{\bf k}_m}=\sqrt{\Big(h+\bar{\rho}\Big)
\Big(h+\bar{\rho}+\bar{\Delta}\Big)}  , 
\end{eqnarray} 
where we also used the fact that the minimum is located along the $z$-axis, so $\theta_{{\bf k}_m}=0$.
The minimum condition $\partial E_{\bf k}/\partial {\bf k}=0$ yields:
\begin{eqnarray}
\label{Emin}
h+\bar{\rho}+\frac{\bar{\Delta}}{4}= h\cdot \frac{\bar{\Delta}}{4\bar{\rho}}\, ,
\end{eqnarray} 
which, in the limit of both reduced exchange and dipolar energy scales being small 
compared to the field, $\bar{\rho}, \bar{\Delta}\ll h$, 
yields a straightforward relation between the two:
\begin{eqnarray}
\label{D_4rho}
\bar{\Delta}=4\bar{\rho}\, .
\end{eqnarray} 
This relation 
gives the field-independent answer for  $|{\bf k}_m|d$:
\begin{eqnarray}
\label{k_m}
|{\bf k}_m|d=\left(\frac{\Delta d^2}{4\rho}\right)^{1/3}\, ,
\end{eqnarray} 
which justifies the above approximation $\bar{\rho}, \bar{\Delta}\ll h$ for large enough $d$ because:
\begin{eqnarray}
\label{D_4rho_1}
\bar{\Delta}=4\bar{\rho}=\left(\frac{4\rho\Delta^2}{d^2}\right)^{1/3}\, .
\end{eqnarray} 
Having the dipolar and exchange scales related by Eq.~(\ref{D_4rho}), 
we can substitute it into Eq.~(\ref{Ekapp}) for $E_{\rm min}$ and use it 
to solve the decay threshold condition $2E_{\rm min}=E_0$ with $E_0=\sqrt{h(h+\Delta)}$. 
This leads to a quadratic equation in the threshold field $h_c(=\mu H_c)$ vs 
$\bar{\Delta}$ (which, in turn,  is a function of $d$ by Eq.~(\ref{D_4rho_1})),
that can be resolved in a compact form:
\begin{eqnarray}
\label{Eq1}
h_{c_{1,2}}=\left(\frac{\Delta}{6}-\bar{\Delta}\right)\pm 
\sqrt{\left(\frac{\Delta}{6}-\bar{\Delta}\right)^2-\frac{5}{12} \, \bar{\Delta}^2}\, ,
\end{eqnarray} 
with two solutions being the upper and the lower threshold fields. They
give the (approximate) threshold boundary 
in Fig.~\ref{Fig3}(b), shown by the dashed line.
From the same solution one can find the ``maximal'' $\bar{\Delta}_m$ that corresponds 
to the lower boundary on $d=d_{\rm min}$ for which  the solution of the threshold condition exists.
This is also the point at which $h_{c_1}=h_{c_2}$ in Eq.~(\ref{Eq1}) and the content of the square-root
is zero, solving for which gives 
\begin{eqnarray}
\label{Eq2}
\bar{\Delta}_m=\bar{\Delta}(d_{\rm min})=\left(\frac{6-\sqrt{15}}{21} \right) \Delta \, .
\end{eqnarray} 
Using Eq.~(\ref{D_4rho_1}) we finally obtain:
\begin{eqnarray}
\label{dmin}
d_{\rm min}=2\left(\frac{21}{6-\sqrt{15}}\right)^{3/2}\sqrt{\frac{\rho}{\Delta}} 
\approx 62.04\,\sqrt{\frac{\rho}{\Delta}}\, .
\end{eqnarray} 
Although the following result can be trivially obtained from the 
solution for $h_{c_1}$ in Eq.~(\ref{Eq1}) taking the $d\rightarrow\infty$ limit ($\bar{\Delta}\rightarrow 0$),
one can derive the large-$d$ limit for $H_{c_1}\rightarrow H_c^{\rm max}$ via an observation that 
at $d\rightarrow\infty$ energy minimum in Eq.~(\ref{Ekapp}) is $E_{\rm min}\approx h$, as 
$\bar{\Delta},\bar{\rho}\rightarrow 0$. This reduces the decay condition to a simple one:
\begin{eqnarray}
\label{simple_decay}
2h_c=\sqrt{h_c(h_c+\Delta)},
\end{eqnarray} 
with two solutions: $h_{c_2}=0$ and $h_{c_1}=\mu H_c^{\rm max}=\Delta/3$, thus relating
the saturated magnetization of the material to the upper limit of the threshold field for decays.

\subsection{threshold field-dependence of $\Gamma_{{\bf k}=0}$}

Since we are interested in the threshold behavior, only spontaneous ($T=0$) decay rate is considered.
Near the threshold,  decays are happening into the magnons in the vicinity of the
 magnon band minima $\pm{\bf q}_{m}=(0,\pm q_{m})$. 
 Expanding in  $({\bf q}-{\bf q}_{m})$ near the minima gives:
\begin{eqnarray}
\label{G1}
&&\Gamma_{{\bf k}=0}=\pi\sum_{\bf q} \big|V^{(3)}_{0;{\bf q}, -{\bf q}}\big|^2
\,\delta\left(E_0-2E_{\bf q}\right) \\
&&= \frac{V_0^2}{2\pi}\int_{-\pi}^\pi dq_y \int_{0}^\pi dq_z \sin^2 2\theta_{\bf q}\,
\delta\left(\Delta E-\frac{\Delta{\bf q}^2}{m}\right)\, ,\nonumber
\end{eqnarray} 
where $\Delta{\bf q}=({\bf q}-{\bf q}_{m})$, $\Delta E=(E_0-2E_{\rm min})$ and  
$m$ is the effective magnon mass. Rewriting $\sin 2\theta_{\bf q}$ as $2q_yq_z/{\bf q}^2$ 
and shifting integration in $q_z$ by $q_{m}$ yields
\begin{eqnarray}
\label{G2}
\Gamma_{{\bf k}=0}=\frac{2V_0^2}{\pi |{\bf q}_{m}|^2}\int q_y^2 \, dq_y \int d\bar{q}_z \, 
\delta\left(\Delta E-\frac{\bar{\bf q}^2}{m}\right)\, ,
\end{eqnarray} 
where $\bar{q}_z =q_z-q_{m}$, $\bar{\bf q}^2=\bar{q}_z^2+q_y^2$, and  
in approximating ${\bf q}^2\approx {\bf q}_{m}^2$ the smallness
of $\bar{q}_z$'s and $q_y$'s in comparison with $|{\bf q}_{m}|$ was used. It finally leads to:
\begin{eqnarray}
\label{G3}
\Gamma_{{\bf k}=0}= \frac{V_0^2m^2}{|{\bf q}_{m}|^2}\,\big|E_0-2E_{\rm min}\big|\, .
\end{eqnarray} 
Since $|E_0\!-\!2E_{\rm min}|\propto |H_c-H|$, this yields $\Gamma_{{\bf k}=0}\propto\Delta H$.

At finite $T\gg E_{\rm min}$, the decay rate near the threshold is
\begin{eqnarray}
\label{G3T}
\Gamma_{{\bf k}=0}= \frac{2T^2}{E_{\rm min}^2}\,\frac{V_0^2m^2}{|{\bf q}_{m}|^2}\,\big|E_0-2E_{\rm min}\big|\, .
\end{eqnarray} 

For the 3D case, when the 2D bands merge into a continuum, the above consideration results in:
\begin{eqnarray}
\label{G4}
\Gamma_{{\bf k}=0}^{3D}\propto \frac{V_0^2}{|{\bf q}_{m}|^2}\int q_\perp^3 \, dq_\perp \, d\bar{q}_z \, 
\delta\left(\Delta E-\frac{\bar{\bf q}^2}{m}\right)\, ,
\end{eqnarray} 
where $\bar{\bf q}^2=\bar{q}_z^2+q_\perp^2$ now and possible mass anisotropies, while neglected,
are not going to change the result qualitatively:
\begin{eqnarray}
\label{G5}
\Gamma^{3D}_{{\bf k}=0}\propto \frac{V_0^2m^2}{|{\bf q}_{m}|^2}\,\big|E_0-2E_{\rm min}\big|^{3/2}\, ,
\end{eqnarray} 
and thus, $\Gamma^{3D}_{{\bf k}=0}\propto \Delta H^{3/2}$.

\begin{figure}[t]
\includegraphics[width=0.99\columnwidth]{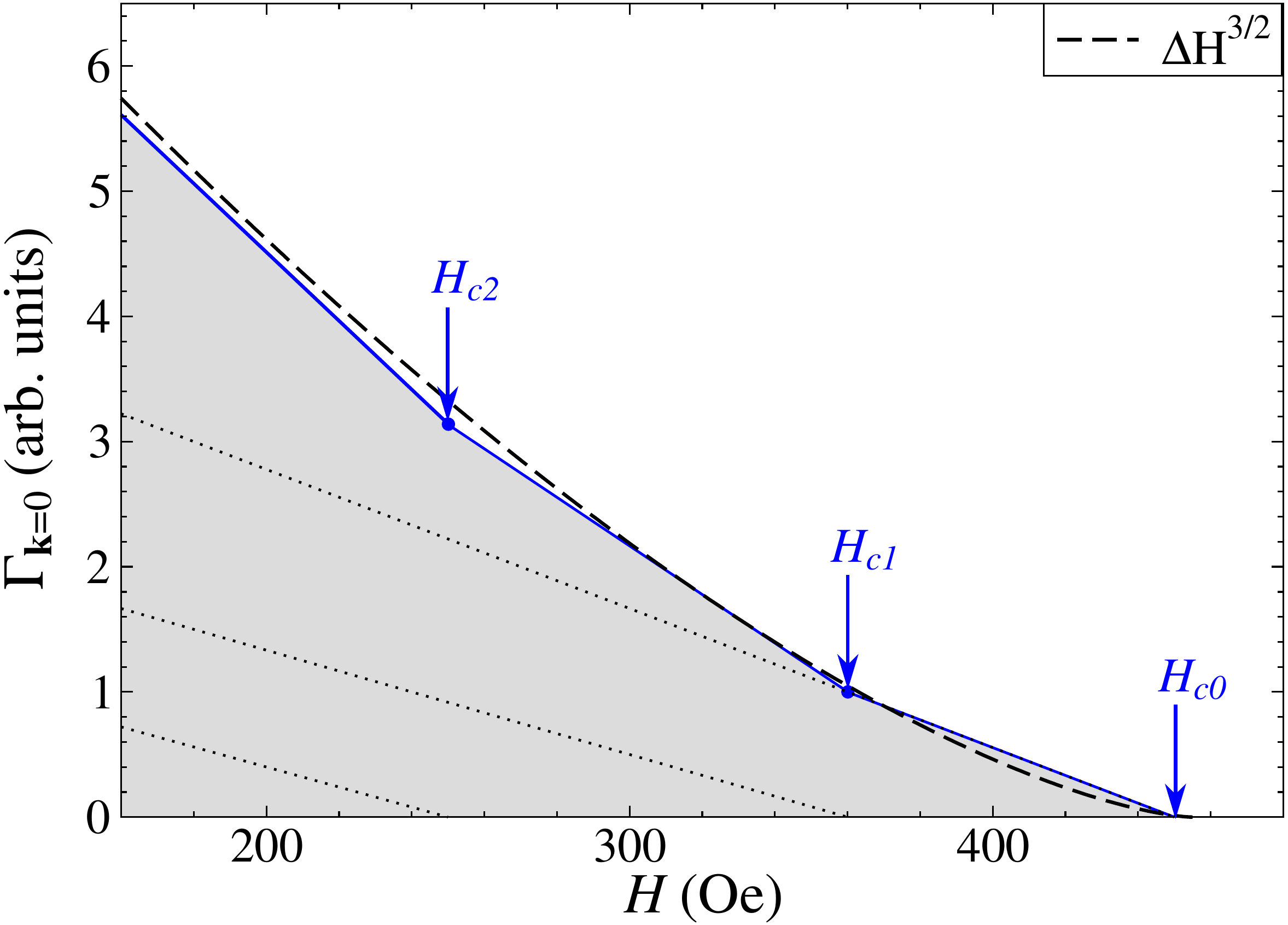}
\caption{(Color online)  A sketch of $\Gamma_{{\bf k}=0}$ vs $H$
for the case of multiple magnon bands. Dotted lines 
show linear in $|H_{c_i}\!-\!H|$ contributions of each band, 
solid line with the shading 
is the total effect, dashed line is the 3D limit $\propto \Delta H^{3/2}$.} 
\label{Fig1supp}
\end{figure} 

\end{document}